\documentclass[aps,a4paper,preprint,longbibliography]{revtex4}%
\usepackage{graphicx}
\usepackage{amsmath}
\usepackage{amsfonts}
\usepackage{amssymb}%
\providecommand{\U}[1]{\protect\rule{.1in}{.1in}}
\begin{document}
\title{Continuous generation of confined bubbles: viscous effect on the
gravito-capillary pinch off}
\author{Haruka Hitomi$^{1}$ and Ko Okumura$^{1\ast}$}
\affiliation{Physics Department and Soft Matter Center, Ochanomizu University, 2-1-1
Ohtsuka, Bunkyo-ku, Tokyo 112-8610, Japan}
\date{\today}

\begin{abstract}
We investigate continuous generation of bubbles from a bath of air in viscous
liquid in a confined geometry. In our original setup, bubbles are
spontaneously generated by virtue of buoyancy and a gate placed in the cell:
the gate acts like an inverted funnel trapping air beneath it before
continuously generating bubbles at the tip. The dynamics is characterized by
the bubble-formation period and the bubble size as a function of the amount of
air under the gate. By analyzing the data obtained for various parameters, we
clearly identified that the dynamics of the bubble formation is governed by
dissipation in the viscous fluid beneath the trapped air balanced by a
gravitational energy change due to buoyancy, after examining numerous
possibilities of dissipation, which demonstrates the potential of scaling
analysis even in complex cases. Furthermore, we uncover a novel type of
pinch-off condition, which convincingly explains the bubble size: in the
present case viscosity plays a vital role, different from the conventional
mechanism of Tate, in which gravity competes with capillarity, revealing a
general mechanism of pinching-off at low Reynolds number. Accordingly, the
present study significantly and fundamentally advances our knowledge of
generation and pinch-off of bubbles, with the results relevant for a wide
variety of applications in many fields. In particular, the present study
demonstrates a promising avenue in microfluidics for understanding physical
principles by scaling up the system, without losing the characteristics of the
flow at low Reynolds numbers.

\end{abstract}
\maketitle

Drop and bubble formation at the end of a tube has been the subject of active
investigations for a long time not only from fundamental but also from
applicational interests in drop and bubbles \cite{DynamicsDroplets}, which
includes oil recovery \cite{HeleShawPetroleum2010}, soft electronics
\cite{babatain2024droplets}, cell therapy \cite{Yasuga2021} and energy
harvesting \cite{xu2020droplet}. As early as in 1864, Tate discussed a pendant
drop at the tip of a tube starts falling when its weight surpasses the
capillary force supporting the weight \cite{Tate}, which remains a fundamental
method to measure surface tension \cite{vinet1993surface,berry2015measurement}%
. A technically more sophisticated modern version of Tate, the continuous
generation of droplets becomes increasingly important in microfluidics
\cite{umbanhowar2000monodisperse,Stone2003,Weitz2007dripping,christopher2007microfluidic,DropletMicrofluids2008,zhu2017passive}
due to recent demand for the manipulation of small amounts of liquids in
various fields such as medicine, biochemistry, and pharmaceutical industries.
However, basic physical principles governing the dynamics of the droplets
formation at small scales and/or at low Reynolds numbers have yet to be
elucidated. One of the difficulties in tackling this problem in microfluidics
is the smallness of the system. One possible strategy to cope with this
difficulty might be to use highly viscous liquid on centimeter scale in
confined space. By virtue of this strategy, we have unveiled a number of
governing principles regarding drop and bubble dynamics in the form of scaling
laws, focusing on viscous friction
\cite{EriSoftMat2011,yahashi2016,murano2020rising,tanaka2023viscous},
coalescence \cite{EriOkumura2010,YokotaPNAS2011,koga2022inertial}, breakup
\cite{nakazato2018self,nakazato2022air}, and bursting
\cite{murano2018bursting}. In this study, we focus on the continuous formation
of bubbles on centimeter scale in a confined geometry, which is much more
directly relevant for microfluidics, to reveal physical principles governing
the dynamics in the form of scaling laws. Using an original setup, we
successfully obtained scaling laws through a clear data collapse with
elucidating physical pictures behind the simple laws.

Salient features of the present study are as follows. (1) To provide an
example of emergence of scaling laws from numerous possibilities of viscous
dissipations; in other words, we have obtained a remarkably simple physics
from a seemingly complex problem, demonstrating the power of scaling analysis.
(2) To provide a condition of breakup in which viscous effect is crucial in
addition to the conventional Tate's mechanism of the balance of gravity and
capillarity, revealing a novel and general mechanism of pinching-off. (3) To
provide an example, in which physical principles relevant for microfluidics
can be effectively elucidated by using a system on centimeter scale without
losing the main character of the flow at low Reynolds numbers. The present
results not only advance fundamental knowledge on drops and bubbles but also
provide guiding principles relevant for numerous applications at low Reynolds
numbers in various fields such as microfluidics and oil industry.

\begin{figure}[h]
\includegraphics[width=\textwidth]{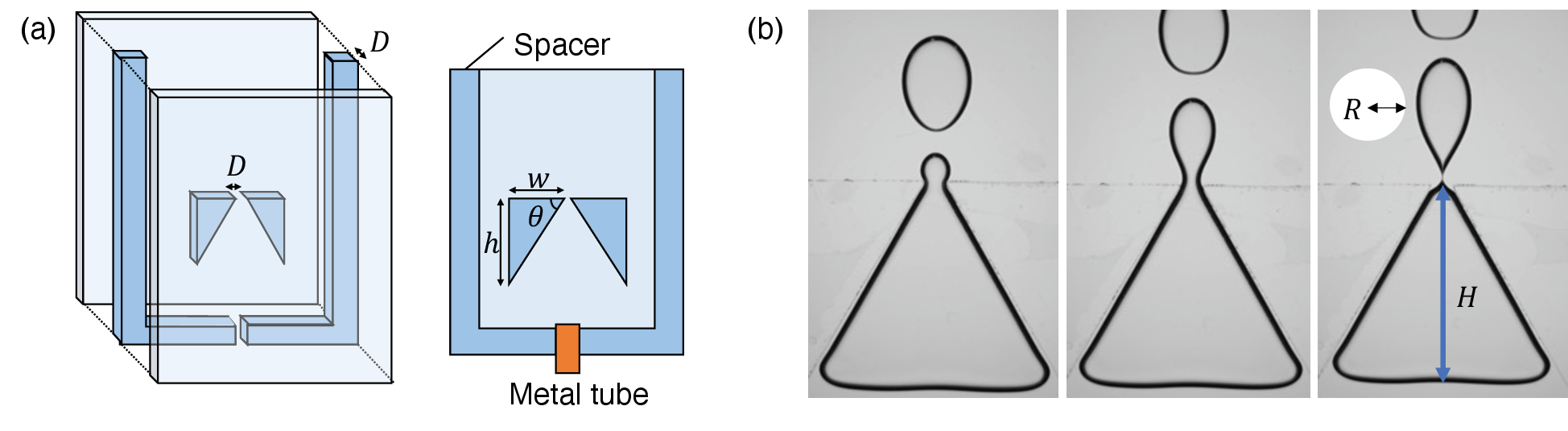}\caption{(a) Experimental setup
with a gate. (b) Continuous generation of bubbles observed in the cell for
$(D,\theta,\nu)=(2,60,30)$ in mm, deg., and St, respectively. The radius $R$
of the white circle having the same area with the bubble on the right
characterizes the size of the bubble.}%
\label{Fig1}%
\end{figure}

\begin{figure}[ptb]
\includegraphics[width=\textwidth]{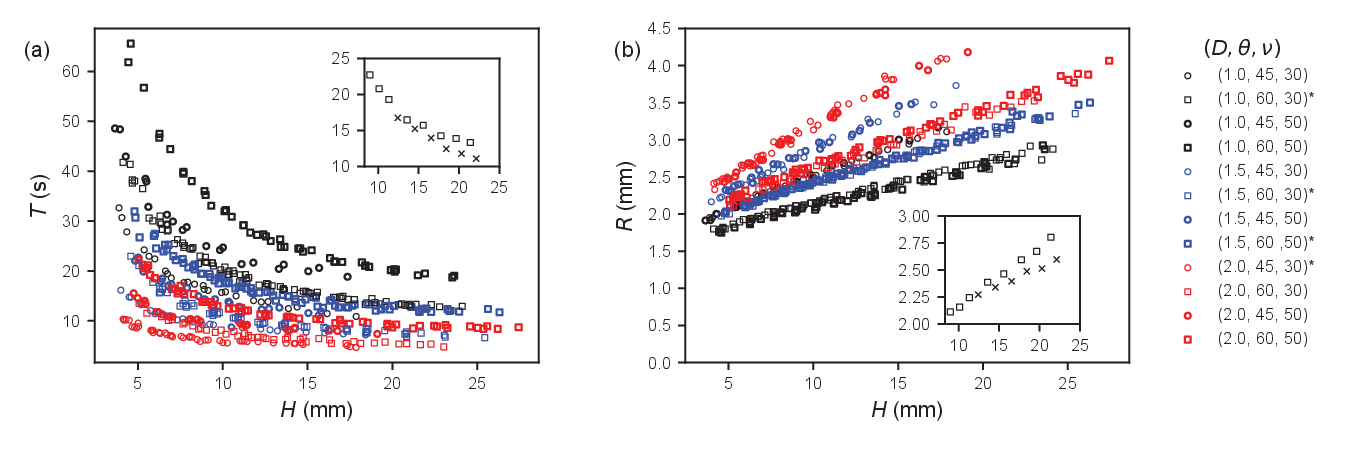} \caption{(a) $T$ vs $H$ and (b)
$R$ vs $H$ on linear scales. The insets demonstrate the existence of the
region in which $T$ and $R$ bifurcate (or oscillate with $H$). See the text
for further details. $D$ and $\nu$ in the legend are given in mm and St,
respectively (note that $30$ St = 3000 cSt).}%
\label{Fig2}%
\end{figure}

In this experiment, we fabricate a thin cell sometimes called a Hele-Shaw cell
(of thickness $D=1.0$ to 2.0 mm, width 15 cm and height 20 cm) equipped with a
gate (of width $D$ and angle $\theta=45$ to 60 deg.) and fill the cell with a
viscous liquid (of kinematic viscosity $\nu=3000$ to 5000 cSt), as in Fig.
\ref{Fig1} (a). We inject air with a syringe through a brass tube (of inner
radius 3.8 mm) at the bottom of the cell to observe a squashed chunk of air
rising in the viscous liquid, which is trapped for a while under the gate with
reducing its mass as a result of continuously generating bubbles, as in Fig.
\ref{Fig1} (b). To characterize the dynamics, we measure the period of
generation $T$ and the characteristic size $R$ ($\gg D$) of the bubble seen
from front as a function of the height $H$ of the air trapped under the gate
[see the rightmost photo of Fig. \ref{Fig1} (b)].

The density $\rho$ and surface tension $\gamma$ of the viscous liquid
[polydimethylsiloxane (PDMS)] are $970$ to $980$ kg/m$^{3}$ and $\gamma=20$
mN/m, respectively. To prevent cell deformation due to capillary adhesion, we
use 5mm-thick acrylic plates for $D=1.0$, 1.5 mm and 2.0 mm-thick acrylic
plates for $D=2.0$ mm. In fact, the cell thickness $D$ is precisely determined
using the laser distance sensor (ZS-HLD2, Omron) and the precise value is used
in the analysis, although, for simplicity, $D$ is represented by approximate
values (1.0, 1.5, or 2.0 mm) as above (differences are within 3\%).

Since the bubble has a tear-drop shape (of area $A$) as seen in Fig.
\ref{Fig1} (b), the characteristic size $R$ is defined through the relation
$A=\pi R^{2}$. The period of generation $T$ for a bubble is defined as the
time difference between the moment of pinch-off the bubble of our focus and
that of the previous bubble. Similarly, $H$ for a bubble of our focus is
defined as the height at the moment of pinch-off of the previous
bubble.\newline

In Fig. \ref{Fig2} (a) and (b), we respectively show $T$ and $R$ as a function
of $H$. The data marked with an asterisk (*) in the legend of Fig. 2 contain
data obtained on different days, where cells were refabricated each day (those
without asterisk are obtained within two hours using the same cell). We
observe that the marked data sets match well with the other data, which
demonstrates a reasonably good reproducibility of the experiment.

The insets shows that $T$ and $R$ as a function of $H$ start bifurcating (or
oscillating with $H$) as $H$ decreases, where we analyze only the data on the
upper branch (we do not use the data represented by the cross mark in the
following). The reason of the bifurcation or oscillation is as follows. Due to
the continuous bubble generation from air trapped under the gate, the volume
of air under the gate (and thus $H$) keep decreasing, and the period of
generation of bubbles $T$ decreases with time (and thus with decrease in $H$).
This implies that the distance between created bubbles becomes short with
time. As a result, at certain point, the upwards flow caused by a bubble just
created could start to affect the creation of the next bubble. If the $n$th
bubble drags the $(n+1)$th bubble, which results in $T$ and $R$ in the lower
branch (represented by the cross mark), then the $(n+2)$th bubble is no longer
affected by the $(n+1)$th bubble. However, the $(n+2)$th bubble does drag the
$(n+3)$th bubble. In this way, we observe the alternate bifurcation (or
oscillation with $H$), where the $(n+2m)$th bubbles (with $m=0,1,2,\ldots$)
are of our focus because the creation of them is completed without the drag
effect of the previous bubble.

\begin{figure}[ptb]
\includegraphics[width=\textwidth]{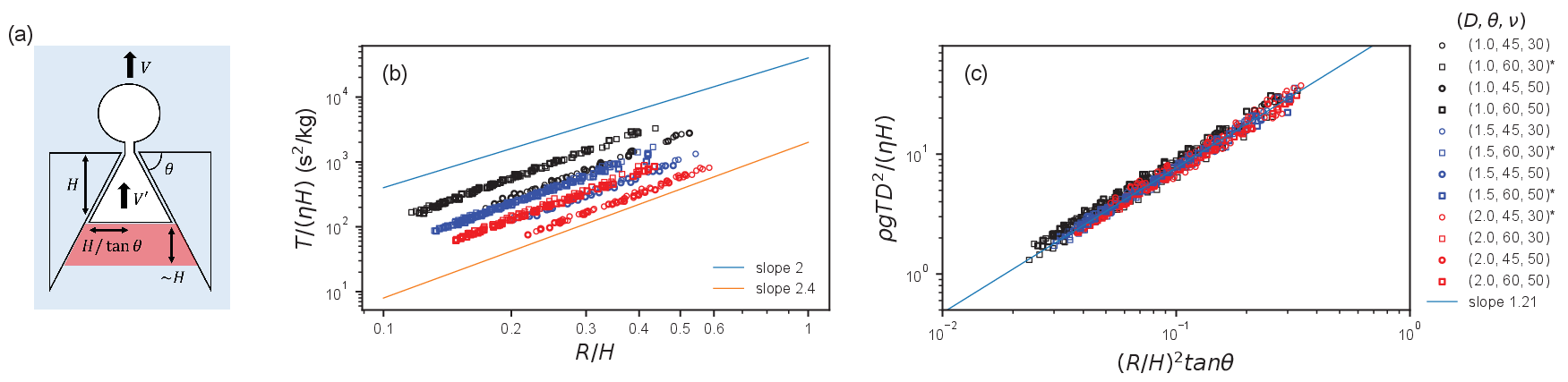} \caption{(a) Physical picture for
the dynamics: The dissipation in the liquid beneath the triangular air pocket
trapped under the gate balances the gravitational energy change due to
buoyancy. (b) $T/(\eta H)$ vs $(R/H)$ on log-log scales, confirming Eq.
(\ref{eq1}). (c) $\rho gTD^{2}/(\eta H)$ vs $(R/H)^{2}\tan\theta$ on log-log
scales, confirming the physical picture in (a), i.e., Eq. (\ref{eq2}).}%
\label{Fig3}%
\end{figure}

Figure \ref{Fig3} explains the relation between $T$ as a function of $H$ with
the illustration in (a) summarizing the physical picture: the dynamics is
determined by the balance between the gravitational energy change due to
buoyancy and viscous dissipation in the viscous liquid beneath the large
triangular air pocket trapped under the gate.

To justify this explanation, we first confirm that the plot in Fig. 3 (b)
clearly support the relation
\begin{equation}
T/(\eta H)=k(D,\theta)(R/H)^{2\alpha} \label{eq1}%
\end{equation}
with $\alpha\simeq1.2$, where the coefficient $k$ is dependent on $D$ and
$\theta$. This is understood as follows. In the plot, colors and symbols
respectively distinguish $D$ and $\theta$, while $\nu$ is distinguished by the
thickness of the outline of a symbol. We observe in the plot the data with the
same color and symbol collapse onto a straight line, while if the color or/and
symbol are different the corresponding data collapse onto a shifted straight
line with the same slope. In this observation, the universal slope corresponds
to $2\alpha$ in Eq. (\ref{eq1}) and the shift reflects the dependence of $k$
on $D$ and $\theta$. If we further note that the data with the same color and
symbol collapse onto the same straight line even if the thickness are
different, we can understand $k$ is independent from $\nu$. In this way, we
can confirm Eq. (\ref{eq1}) in Fig. 3 (b).

Second, we consider numerous possibilities of dissipation balanced with the
energy change due to buoyancy, as explained in detail in Fig. \ref{Fig5}
below. In this consideration, Eq. (\ref{eq1}) is very helpful, since if a
possibility of dissipation does not lead to a result that is consistent with
Eq. (\ref{eq1}) we can exclude the possibility as not playing a dominant role.
In other words, Eq. (\ref{eq1}) gives a restriction on the possibility.
Remarkably, this restriction is very sever and we find that only one single
possibility is consistent with Eq. (\ref{eq1}).

This only possibility is illustrated in Fig. 3 (a), where the large triangular
air pocket under the gates is moving upwards with a velocity $V^{\prime}$.
This velocity can be estimated from a volume conservation as $V^{\prime
}T(2H/\tan\theta)D=\pi R^{2}D$, which can be expressed as%
\begin{equation}
V\sim V^{\prime}H/(R\tan\theta)\text{ with }R\sim VT, \label{eqa}%
\end{equation}
and necessarily induces a velocity gradient $\sim V^{\prime}/D$ inside the
viscous fluid in the vicinity of the bottom of the air pocket in a volume
$\sim DH^{2}/\tan\theta$, comparable to the volume of the triangular air
pocket, as indicated in the illustration. We balance the dissipation energy
during the time $T$ (i.e. during the creation of one bubble) with the
corresponding change in gravitational energy: $\rho gR^{2}DH\simeq
T\eta(V^{\prime}/D)^{2}(H^{2}/\tan\theta)D$. From the volume conservation and
energy balance, we obtain%
\begin{equation}
\rho gTD^{2}/(\eta H)\simeq(R/H)^{2}\tan\theta, \label{eq2}%
\end{equation}
which is in accord with Eq. (\ref{eq1}): $k(D,\theta)\simeq\tan\theta/(\rho
gD^{2})$ and $\alpha\simeq1$.

This relation is convincingly confirmed by a clear collapse of data shown in
Fig. \ref{Fig3} (c) without any fitting parameters, although the agreement is
not perfect. The slope ($\alpha$) of the straight line in (c) obtained by
numerical fitting is $\alpha=1.21\pm0.04$, which is slightly larger than the
expected value of $\alpha=1$.

Strictly speaking, the bubble and air pocket are both covered with lubricating
films (of thickness $h$) of the viscous fluid, which implies that there is no
direct contact of air with the surface of the cell plates. However, we expect
this thickness is very thin and for simplicity we assumed $D\gg h$, in the
above and the ensuing discussion (a quantitative justification is also
possible, as given below).

\begin{figure}[ptb]
\includegraphics[width=\textwidth]{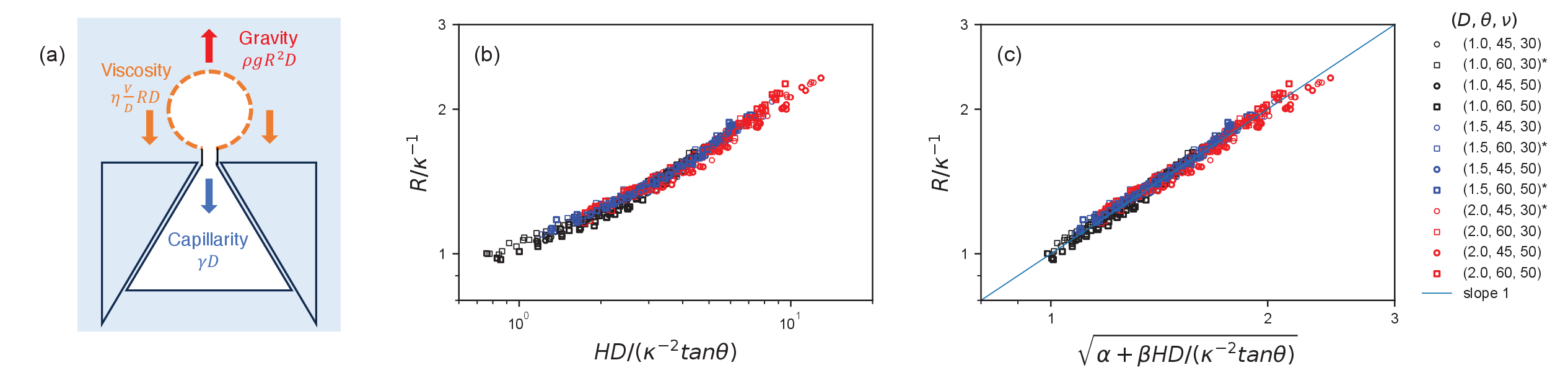} \caption{(a) Physical picture for
pinch-off: Buoyancy is opposed by viscosity in addition to capillarity. (b)
$R/\kappa^{-1}$ vs $HD/\tan\theta$ on log-linear scales, confirming Eq.
(\ref{eq4}) without any fitting parameter. (c) $R/\kappa^{-1}$ vs
$\sqrt{\alpha+\beta HD/(\kappa^{-2}\tan\theta)}$ on linear scales,
demonstrating an excellent agreement, with using the result of fitting for
$\alpha$ and $\beta$ specified in the text.}%
\label{Fig4}%
\end{figure}

Figure \ref{Fig4} explains the relation between $R$ as a function of $H$ with
the illustration in (a) summarizing a surprising physical picture: buoyancy
opposed not only by capillarity but also by viscosity determines the condition
of pinch-off, different from Tate's law. If we consider a natural form of
viscous stress $\eta V/D$ acting on the circumference of the disk-shaped
bubble whose area scales as $RD$, we obtain a force balance%
\begin{equation}
\rho gR^{2}D=\alpha\gamma D+\beta\eta VR\label{eq3}%
\end{equation}
with dimensionless coefficients $\alpha$ and $\beta$. To remove $V$ from this
equation, we use an equation for the bubble velocity $R\simeq VT$ and Eq.
(\ref{eq2}) for $T$, we arrive at the following relation based on the
unexpected pinch-off condition with the capillary length $\kappa^{-1}%
=\sqrt{\gamma/(\rho g)}$:%
\begin{equation}
R/\kappa^{-1}=\sqrt{\alpha+\beta HD/(\kappa^{-2}\tan\theta)}\label{eq4}%
\end{equation}

This equation reveals that the dimensionless quantity $R/\kappa^{-1}$ should
be a function of a dimensionless quantity $HD/(\kappa^{-2}\tan\theta)$, which
is convincingly confirmed in Fig. \ref{Fig4} (b) without any fitting
parameters. We further use Eq. (\ref{eq4}) to fit the data to obtain
$\alpha=0.665\pm0.04$ and $\beta=0.411\pm0.02$ by taking averages of numerical
fitting for each parameter. An excellent agreement between this result of
fitting and the data is shown in Fig. \ref{Fig4} (c).

The collapse seen in Fig. \ref{Fig4} (b), or Eq. (\ref{eq4}), suggests that
$R$ is independent from $\eta$. This is consistent with the balance in Eq.
(\ref{eq3}), in which the last term describing the shear stress is in fact
independent of $\eta$, as encountered in Poiseuille flow in a pipe: The shear
stress $\eta VR\sim\eta R^{2}/T$ is independent from $\eta$, because $T$ is
proportional to $\eta$, as suggested by the collapse in Fig. 3 (c), or Eq.
(\ref{eq2}).

\begin{figure}[ptb]
\includegraphics[width=0.6\textwidth]{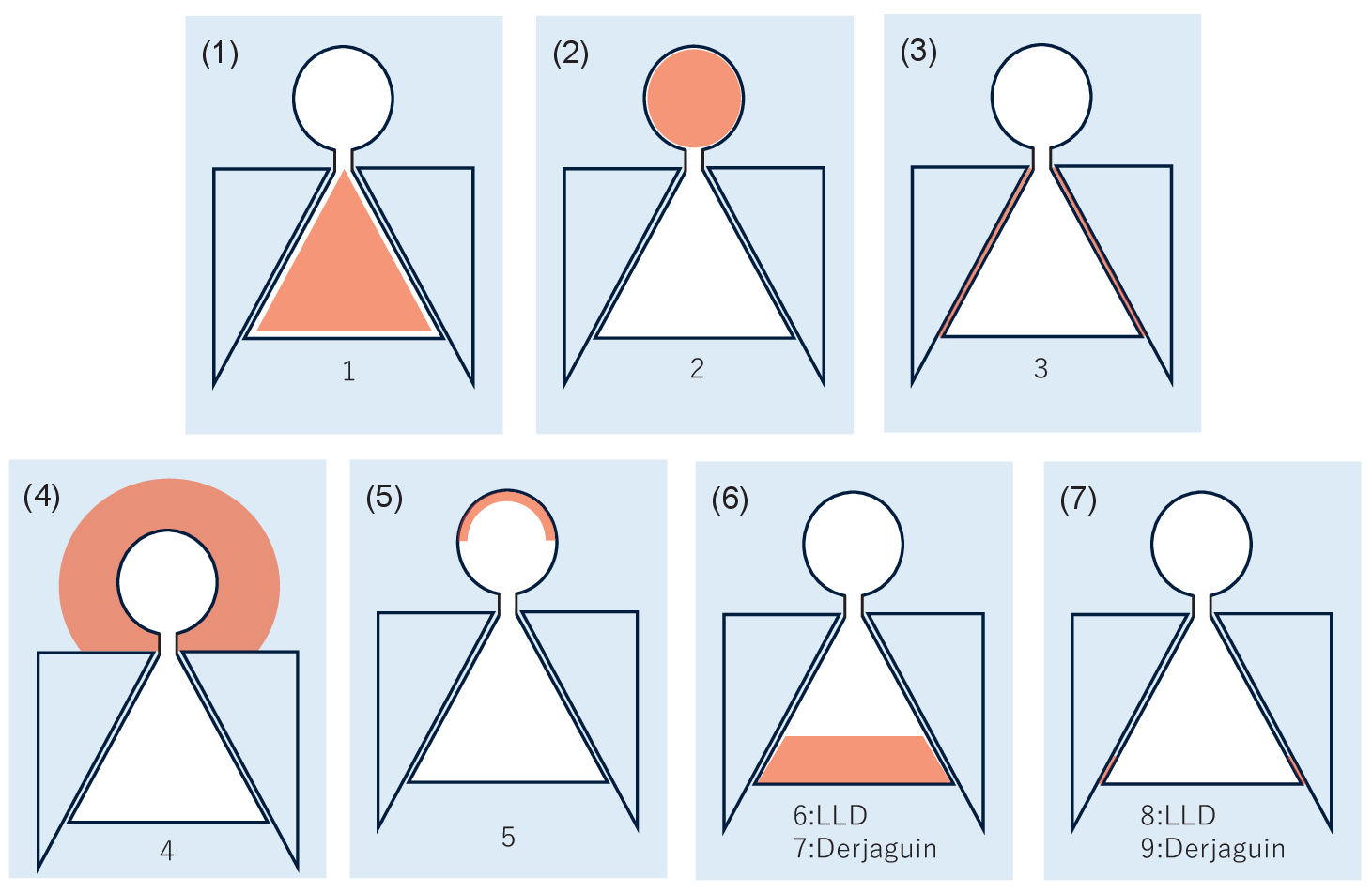} \caption{Numerous
possibilities of viscous dissipation considered in the present study. The red
region indicates the place in which a velocity gradient is developed.}%
\label{Fig5}%
\end{figure}

Although we focus on the conditions at the pinch off in the above scaling
arguments, we can discuss some dynamical picture leading to the pinch off.
Just after a pinch off, the triangular air pocket of height $H$ starts to push
a new bubble through the orifice, which is possible because the capillary
force at the orifice $\simeq\gamma D$ is smaller than the buoyancy $\simeq\rho
gDH^{2}/\tan\theta$, i.e., $\kappa^{-1}<H$. Then the liquid below the bubble
starts to move and viscous dissipation characterized by a velocity gradient
$V^{\prime}/D$ starts building up beneath the triangular air pocket of height
$H$ to reach a force balance $\rho gDH^{2}/\tan\theta$ $\simeq$ $\eta
(V^{\prime}/D)H^{2}/\tan\theta$ at the velocity
\begin{equation}
V^{\prime}\simeq\rho gD^{2}/\eta, \label{eqa2}%
\end{equation}
which is independent of $H$. From this expression for $V^{\prime}$ and the
volume conservation in Eq. (\ref{eqa}), we recover Eq. (\ref{eq2}) previously
obtained by an energy balance. To deepen the dynamical picture given above, we
consider the governing forces acting on the growing bubble: the upward
buoyancy $\sim\rho gR^{2}D$ is opposed by the downward forces of viscous and
capillary origins, $\sim$ $\eta(V/D)RD$ and $\sim\gamma D$. Dynamically, while
the last capillary force has a geometrical maximum, the upward buoyancy ($\sim
R$) increases in magnitude and the downward viscous force, which can be
expressed as $\rho gD^{2}H$ using Eqs. (\ref{eqa}) and (\ref{eqa2}),
decreases. This dynamical balance implies that the buoyancy eventually becomes
dominant leading to the pinch off, at which Eq. (\ref{eq3}) is satisfied.
Furthermore, the viscous force acting on the bubble, which eventually appears
in Eq. (\ref{eq3}), corresponds to a viscous dissipation per time around the
bubble characterized by the velocity gradient $V/D$ in a volume $RD^{2}$ (see
Appendix \ref{app2}). This dissipation can be shown to be smaller than the
dissipation beneath the triangular pocket considered in deriving Eq.
(\ref{eq2}), from Eqs. (\ref{eqa}) and (\ref{eqa2}), which supports the
derivation of Eq. (\ref{eq2}).

Throughout the present study, we ignored the effect of inertia, which is
justified as follows. We can estimate the upper bound for Reynolds number by
$Re=$ $\eta VL/\eta$ once a relevant characteristic length scale $L$ is
identified. Judging from Eqs. (\ref{eq2}) and (\ref{eq4}), it is natural to
consider that $L$ would scale as $\kappa^{-1}$, which is comparable with $D$.
Then, considering the range of parameters in the present study, we confirmed
$Re$ is less than 0.0005 (for $L=1.8$ mm), which means $Re\ll1$, as we
assumed. In the above, we used $V$ instead of $V^{\prime}$, which implies $Re$
could be overestimated because $V>V^{\prime}$. This guarantees that the above
conclusion $Re\ll1$ is valid even if this issue is taken into account.

We examine the validity of the assumption $D\gg h$. The thickness $h$ of the
lubricating film can be estimated by the theory of Landau, Levich, and
Derjaguin (LLD) \cite{LandauLevich,Derjaguin1943} or that of Derjaguin
\cite{Derjaguin1943,derjaguin1993thickness}: $h$ can be respectively estimated
as $h_{LLD}=0.94\kappa^{-1}Ca^{2/3}$ and $h_{D}=\kappa^{-1}Ca^{1/2}$ with the
capillary number $Ca=\gamma U/\eta$ for a characteristic velocity $U$, which
suggests the use of $V$ for $U$ could give an overestimate for $Ca$. In the
present study, if we estimate $Ca$ with $U=V$, $Ca$ is in the range from
0.00661 to 0.146, with the average 0.045 and the standard deviation 0.03, to
confirm $Ca<1$. Note that this conclusion $Ca<1$ is valid even if we use
$U=V^{\prime}$ since $V>V^{\prime}$. Based on the range of $Ca$ thus obtained,
the theory of Derjaguin predicts $h/D$ to be in the range from 0.12 to 0.19
with the average 0.16 and the standard deviation 0.01, to reasonably well
confirm $h/D\ll1$. This conclusion $h/D\ll1$ does not change even if we use
$U=V^{\prime}$ and/or LLD theory because both replacements could only lead to
a smaller value of $h/D$ (Note here $h_{D}\sim Ca^{1/2}$ $>$ $h_{LLD}\sim
Ca^{2/3}$ if $Ca<1$): the above estimate could be an overestimate for these
two aspects (If $\kappa^{-1}\gtrsim D$ as in the present study, $h_{LLD}$ and
$h_{D}$ could scale not with $\kappa^{-1}$ but rather with $D$; e.g.,
$h_{D}\sim DCa^{1/2}$; the above estimate based on $h_{D}=\kappa^{-1}Ca^{1/2}$
could be an overestimate even in this respect).

As announced previously, we considered various possibilities for dissipation
other than the one considered in the above, as specified in Fig. \ref{Fig5}
(1) to (7), where the region in which the velocity gradient is developed is
suggested in red. In Fig. \ref{Fig5} (1) to (3), the dissipation in thin films
is indicated. However, because of a low viscosity of air, even though air is
moving, the air cannot drag the liquid surface effectively to induce a
significant velocity gradient. For this reason, we exclude these three
possibilities. In Fig. \ref{Fig5} (4), a velocity gradient $\sim V/D$ is
developed around the bubble in a volume $\sim R^{2}D$ or $\sim RD^{2}$ (see
Appendix \ref{app2}). However, this dissipation does not give any $\theta$
dependence, which is inconsistent with Eq. (\ref{eq1}), and thus is excluded.
In Fig. \ref{Fig5} (5) to (7), velocity gradients developed inside the dynamic
meniscus in thin films are considered. However, the case of Fig. \ref{Fig5}
(5) again fails to give any $\theta$ dependence, and thus is excluded. The
remaining cases of Fig. \ref{Fig5} (6) and (7) are also excluded because we
again fail to obtain the form consistent with Eq. (\ref{eq1}) even though we
have some $\theta$ dependence and even though we consider the two
possibilities of $h=h_{LLD}$ and $h=h_{D}$ in each case, as shown in Appendix
\ref{app1}.

In this way, although we tested nine cases as suggested in Fig. \ref{Fig5},
all were excluded: the only possibility is the one leading to Eq. (\ref{eq2}).
In other words, the present case is rather complex in that there are many
possibilities for the region in which dissipation could occur. Despite this
complexity, a governing dissipation is singled out from numerous possibilities
to results in simple and clear scaling laws. In this sense, the present study
is a remarkable example of a simple physics emerging from complexity.

In addition, the effect of viscosity on the Tate's condition of pinch-off
uncovered in the present study should be a general mechanism to be considered
in many other cases in microfluidics or/and at low Reynolds numbers. Together
with this, the present study provides a clear and fundamental physical
understanding for the dynamics of continuous bubble formation relevant for
numerous applications, advancing and impacting on the field, demonstrating a
system on centimeter scales could be useful.

\section*{Acknowledgments}

We are grateful to an anonymous referee, who gave us deep and insightful
comments, including those on the dynamical picture of the present pinch off.
We thank Mana Iwasaki and Yuka Katsumata for contribution for initial stage of
the present work. This work was supported by JSPS\ KAKENHI Grant Number
JP19H01859 and JP24K00596.


\renewcommand{\thesection}{} \setcounter{section}{0}
\renewcommand{\thesubsection}{A\arabic{subsection}} \setcounter{subsection}{0} \renewcommand{\theequation}{A\arabic{subsection}.\arabic{equation}}

\section{Appendix \label{APP}}


\subsection{Dissipation around the growing bubble\label{app2}}

The present experimental data strongly suggest that the dissipation (per time)
around the growing bubble is characterized by a velocity gradient $\sim V/D$
in a volume $\sim RD^{2}$ (not $\sim R^{2}D$), which can be shown to be
smaller than the dissipation beneath the triangular air pocket as discussed in
the text. This is understood from the shear force considered in Eq.
(\ref{eq3}) $\sim\eta(V/D)RD$, which is transformed into dissipation energy
per time by multiplying a factor $\sim V$: $\eta(V/D)^{2}RD^{2}$. This is in
contrast with the dissipation per time around a rising bubble, which is not
growing but has a fixed volume, in a Hele-Shaw cell, which is characterized by
a velocity gradient $\sim V/D$ in a volume $\sim R^{2}D$ in previous studies
(see \cite{EriSoftMat2011} and references therein). Note that the shear force
considered in Eq. (\ref{eq3}) cannot be interpreted as $\sim\eta(V/R)R^{2}$
because it leads to the dissipation $\sim\eta(V/R)^{2}R^{3}$, which is rather
for a three-dimensional bubble.

\subsection{Scaling arguments for the cases in Fig. \ref{Fig5} (6) and
(7)\label{app1}}

Before going into the specific cases, we slightly extend the theory of LLD and
Derjaguin at the level of scaling laws, for a film of thickness $h$. In the
film, a velocity gradient $\sim U/h$ is developed only in the region of
dynamic meniscus whose length scale is characterized by $l$. The LLD theory is
based on the Stokes equation and matching condition:%
\begin{equation}
\eta\frac{U}{h^{2}}\sim\frac{1}{l}\frac{\gamma}{l_{0}}\text{ and }\frac
{1}{l_{0}}\sim\frac{h}{l^{2}}. \label{b1}%
\end{equation}
In the first relation, the viscous stress balances with the pressure gradient
originating from Laplace's pressure jump $\gamma/l_{0}$. In the second, the
radius of curvature $l^{2}/h$ is matched with the length $l_{0}$, which is the
capillary length $\kappa^{-1}$ in the original theory but here is replaced by
the cell thickness $D$ if $D\ll\kappa^{-1}$, where this replacement implies a
slight extension of the original theory. From Eq. (\ref{b1}), we obtain%

\begin{equation}
l\sim l_{0}(\eta U/\gamma)^{1/3}\text{ and }h\sim l_{0}(\eta U/\gamma)^{2/3},
\label{b2}%
\end{equation}

In Derjaguin's theory, Eq. (\ref{b1}) is replaced by%
\begin{equation}
\eta\frac{U}{h^{2}}\sim\rho g\text{ and }\frac{1}{l_{0}}\sim\frac{h}{l^{2}},
\end{equation}
from which we obtain%
\begin{equation}
l\sim(\kappa^{-1}l_{0})^{1/2}(\eta U/\gamma)^{1/4}\text{ and }h\sim\kappa
^{-1}(\eta U/\gamma)^{1/2}, \label{b4}%
\end{equation}

In Fig. \ref{Fig5} (6), we consider a velocity gradient $\sim U/h$ with
$U=V^{\prime}$ inside the film (thickness $h$) developed in the dynamic
meniscus contacting on the surface of cell plates in a volume $\sim
hlH/\tan\theta$, which leads to the following gravity-viscous energy balance
during the time comparable to $T$:%
\begin{equation}
\rho gR^{2}DH\sim T\eta(V^{\prime}/h)^{2}hlH/\tan\theta\label{a1}%
\end{equation}

If we use the LLD theory, this balance can be reduced in the following form in
which the $l_{0}$ dependences are cancel out, by using Eq. (\ref{b2}) with
$U=V^{\prime}$ together with the volume conservation $V^{\prime}%
T(2H/\tan\theta)D=\pi R^{2}D$:%
\begin{equation}
\left(  \frac{\rho gTD^{2}}{\eta H}\right)  ^{2}\simeq\tan^{2}\theta
\frac{\kappa^{-2}R^{4}D}{H^{7}}\text{,} \label{a2}%
\end{equation}
which is not compatible with Eq. (\ref{eq1}), and thus this case is excluded
from possible candidates.

If we instead use Derjaguin theory, the energy balance in Eq. (\ref{a1})
reduces to the following form, with the aid of Eq. (\ref{b4}) with
$U=V^{\prime}$ together with the same volume conservation:%
\begin{equation}
\left(  \frac{\rho gTD^{2}}{\eta H}\right)  ^{3}\simeq\tan^{3}\theta l_{0}%
^{2}R^{6}D^{2}/H^{10},
\end{equation}
which is again not compatible with Eq. (\ref{eq1}), and thus this case is also excluded.

In Fig. \ref{Fig5} (7), we consider a velocity gradient $\sim U/h$ with
$U=V^{\prime}/\sin\theta$ inside the film (thickness $h$) developed in the
dynamic meniscus contacting on the surface of the gates in a volume $\sim
hlD$, which lead to the following gravity-viscous energy balance during the
time comparable to $T$:%
\begin{equation}
\rho gR^{2}DH\sim T\eta\left(  \frac{V^{\prime}}{h\sin\theta}\right)  ^{2}hlD
\label{c1}%
\end{equation}

If we use the LLD theory, this balance can be reduced in the following form in
which the $l_{0}$ dependences are cancel out, by using Eq. (\ref{b2}) with
$U=V^{\prime}/\sin\theta$ together with the same volume conservation
$V^{\prime}T(2H/\tan\theta)D=\pi R^{2}D$:%
\begin{equation}
\left(  \frac{\rho gTD^{2}}{\eta H}\right)  ^{2}\simeq\frac{\kappa^{-2}%
R^{4}D^{4}}{H^{10}\cos^{5}\theta}\text{,} \label{c2}%
\end{equation}
which is not compatible with Eq. (\ref{eq1}), and thus this case is excluded
from possible candidates.

If we instead use Derjaguin theory, Eq. (\ref{c1}) reduces to the following
form, with the aid of Eq. (\ref{b4}) with $U=V^{\prime}/\sin\theta$ together
with the same volume conservation:%
\begin{equation}
\left(  \frac{\rho gTD^{2}}{\eta H}\right)  ^{3}\simeq\frac{l_{0}^{2}%
R^{6}D^{6}}{H^{14}\cos^{7}\theta},
\end{equation}
which is again not compatible with Eq. (\ref{eq1}), and thus this case is also excluded.

In this way, even though we considered four cases corresponding to Fig.
\ref{Fig5} (6) and (7), all the cases are not in the form of Eq. (\ref{eq1})
and thus are excluded.


\end{document}